\documentclass[12pt,a4paper,notcite,notref]{article}

\usepackage{fancybox,amsmath,epsf}

\def \t#1{\widetilde{#1}}
\def \h#1{\widehat{#1}}

\def \PdE{{P$\Delta$E}}

\usepackage{epsfig}
\usepackage{eucal}
\usepackage{epic}

\newtheorem{defn}{Definition}

\title{Hirota's method and the search for\\
  integrable partial difference equations.\\
1. Equations on a $3\times 3$ stencil}

\author{Jarmo Hietarinta$^1$\footnote{E-mail: jarmo.hietarinta@utu.fi}
  ~ and ~ Da-jun Zhang$^2$\footnote{E-mail: djzhang@staff.shu.edu.cn }
  \\
  {\small\it $^1$Department of Physics and Astronomy,
    University of Turku, FIN-20014 Turku, Finland} \\
  {\small\it $^2$Department of Mathematics, Shanghai University,
    Shanghai 200444, P.R. China}}

\date{\today}

\begin{document}

\maketitle

\begin{abstract} Hirota's bilinear method (``direct method'') has been
  very effective in constructing soliton solutions to many integrable
  equations. The construction of one- and two-soliton solutions is
  possible even for non-integrable bilinear equations, but the
  existence of a generic three-soliton solution imposes severe
  constraints and is in fact equivalent to integrability.  This
  property has been used before in searching for integrable partial
  differential equations (PDE), and in this paper we apply it to two
  dimensional partial difference equations (\PdE) defined on a
  $3\times 3$ stencil. We also discuss how the obtained equations are
  related to projections and limits of the three-dimensional master
  equations of Hirota and Miwa, and find that sometimes a
  singular limit is needed.
\end{abstract}

\section{Introduction}
Integrable dynamics is fundamentally associated with regularity and
predictability, but if the equation is nonlinear this does not imply
that the behavior is simple, quite the contrary.

There are in fact many interesting mathematical properties that are
associated with nonlinear integrable PDE's (soliton equations), such
as the existence of sufficient number of conserved quantities and
symmetries, the ability to write the PDE as the consistency condition
of two linear equations (the Lax pair, with spectral parameter), the
existence of multi-soliton solutions, etc.\cite{genPDE}.  Any of these
characterizing properties can also be used to search for integrable
PDE's and it is remarkable that in the end each criterion gives
essentially the same set of equations.

In recent years the study of integrable partial {\em difference}
equations (P$\Delta$E) has progressed rapidly \cite{SIDE}. The above
integrability indicators have also been applied to P$\Delta$E's, but
in the discrete case there are also other criteria, specific to
discrete equations, such as singularity confinement\cite{sing}, the so
called ``Consistency-Around-the-Cube'' property \cite{CAC}, and slow
growth of complexity of the iterates (algebraic entropy)\cite{ae}.

In this paper we consider the {\em existence of three-soliton
  solutions} for P$\Delta$E's as an indicator of integrability. We use
Hirota's bilinear method, which is well suited for constructing
soliton solutions. In Section 2 we first describe how the method works
in searching for integrable PDE's and then formulate the search
problem for P$\Delta$E's. In Section 3 we give the results for a
particular class of two-dimensional lattice equations, namely those
defined on a $3\times 3$ stencil. It is remarkable that all of the
integrable equations obtained from this class can in fact be described
as projections and limits of the well known three-dimensional bilinear
master-equations or Hirota \cite{Hir81} or Miwa \cite{Miw82}, although
sometimes a singular limit is needed. This is shown in Section 4.

\section{Hirota's direct method}
\subsection{The continuum case}
In 1971 R.~Hirota proposed a direct method for constructing soliton
solutions \cite{Hir71}. A key ingredient in this method is a
transformation to {\em new dependent variables} so that in these new
variables a soliton solution is a finite sum of exponentials.

\subsubsection{Definition}
For the prototypical example of KdV equation
\[
u_t+u_{xxx}+6uu_x=0,
\]
one applies the transformation
\begin{equation*}
u=2\partial_x^2\log F,
\end{equation*}
and after integrating the resulting equation once one gets a bilinear
equation
\begin{equation}\label{KdVbil}
(D_x^4+D_xD_t)F\cdot F=0,
\end{equation}
were the {\em Hirota derivative operator} $D$ is defined by
\begin{eqnarray}
D_x^n\,f\cdot g&=&(\partial_{x_1}-\partial_{x_2})^nf(x_1)g(x_2)
\big |_{x_2=x_1=x}\label{eq:Hder} \\
&\equiv&\partial_y^n\,f(x+y)g(x-y)\big |_{y=0}.\nonumber
\end{eqnarray}
Thus  $D$ acts on a product similarly to
the Leibniz rule, except for a crucial change of sign. In particular
$D_x^n\,f\cdot g=(-1)^nD_x^n\,g\cdot f$.  Equations in which
derivatives appear only through the above $D$-operator are said to be
in {\em Hirota bilinear form}. (For further properties of the $D$
operator, see the book \cite{Hbook} and Appendix I of \cite{HSold}.)

The first step in constructing solitons solutions is therefore to find
the dependent variable transformation that leads to a Hirota bilinear
form.  Unfortunately the bilinearization of a given nonlinear PDE is
not easy. One cannot know before hand how many dependent or even
independent variables one need to properly bilinearize a given
equation.\footnote{In the literature one sometimes finds examples of
  over-bilinearization, in which a multi-linear equation is broken
  down into too many bilinear equations. One case study can be found
  in \cite{GHNO03}.}

\subsubsection{Gauge invariance}
At this point we would like to emphasize that equations written in
Hirota bilinear form can also be characterized by {\em gauge
  invariance}.  Consider the gauge transformation $(f,g,\dots)\mapsto(
e^{\vec x\cdot\vec p}f, e^{\vec x\cdot\vec p}g, \dots)$, then the
condition for gauge invariance is
\begin{equation}  \label{eq:gauge}
  P(\vec D)(e^{\vec x\cdot \vec p}f)\cdot (e^{\vec x\cdot \vec p}g)=
        e^{2\vec x\cdot\vec p}P(\vec D)f\cdot g.
\end{equation}
(Here we use vector notation: $\vec{D}=(D_x,\,D_y,\,D_z,\dots)$, $\vec
x=(x,y,z,\cdots)$, $\vec p=(p_x,p_y,p_z,\cdots)$ with inner product
$\vec x\cdot \vec p=p_x\,x+p_y\, y+\cdots$.)  Equations written in
Hirota bilinear form have this property, and conversely, any quadratic
equation having gauge invariance of the above type can be written in
Hirota bilinear form \cite{GRH94}.

Gauge invariance allows generalizations of the Hirota bilinear form to
other circumstances, such as higher multi-linearity \cite{GRH94} and
discrete equations.

\subsubsection{1SS and 2SS}
The existence of a bilinear form does not guarantee the existence
of $N$-soliton solutions, but one- and two-soliton solutions are often
easy to construct. As an example take the bilinear equation
\begin{equation}\label{eq:generic}
P(\vec D)F\cdot  F=0
\end{equation}
where $P$ is some {\em even} polynomial in $\vec D$ (such as
\eqref{KdVbil} in the KdV case).

In order to start the expansion we need a vacuum or background
solution, which in this case is $F\equiv 1$ (which in the KdV case
corresponds to $u\equiv0$). This works, provided that $P(0)=0$.

Starting with the vacuum $F=1$ one builds a one-soliton solution (1SS)
``perturbatively'' using exponentials of linear functions, which are
called plane wave factors (PWF):
\begin{equation}\label{eq:1SS}
F=1+\epsilon\, e^{\eta_i},\text{ where } \eta_i=\vec x\cdot\vec p_i+\eta_i^0,
\text{ with constant } \eta_i^0.
\end{equation}
Here $\epsilon $ is the formal expansion parameter, and upon
substituting this to \eqref{eq:generic} one finds the constraint that
the parameters $\vec p_i$ must satisfy the {\em dispersion relation}
(DR)
\begin{equation}\label{eq:genDR}
P(\vec p_i)=0.
\end{equation}

Continuing in this manner one finds that the generic equation
\eqref{eq:generic} also has two-soliton solutions of the form
\begin{equation}
F=1+\epsilon\, e^{\eta_1}+\epsilon\, e^{\eta_2}+\epsilon^2\, A_{12}e^{\eta_1+\eta_2}, \label{eq:2SS}
\end{equation}
where
\begin{equation}
A_{ij}=-\frac{P(\vec p_i-\vec p_j)}{P(\vec p_i+\vec p_j)}. \label{eq:A12}
\end{equation}
Here each $\vec p_i$ is restricted by the DR
\eqref{eq:genDR}, that already appeared in constructing the 1SS, but
there are no additional restrictions on them. $A_{ij}$ is called the {\em
  phase factor}.
Note that $A_{ij}=A_{ji}$ due to $P$ being even.

In the above constructions crucial role is played by the minus sign in
the definition \eqref{eq:Hder} of the Hirota derivative, which implies
\begin{equation}
  \label{eq:tail}
  P(\vec D)e^{\vec x\cdot \vec p}\cdot e^{\vec x\cdot {\vec p}\, '}=
        P(\vec p-\vec{p}\,')e^{\vec x\cdot(\vec p+{\vec p}\, ')}.
\end{equation}
In practice this means that the first {\em and} the last terms of
expressions like the 1SS \eqref{eq:1SS} and 2SS \eqref{eq:2SS}
automatically satisfy the equation (if $P(0)=0$).

The above construction shows a level of {\em partial integrability}:
we can have elastic scattering of two solitons, for any dispersion
relation, if the nonlinearity is suitable (namely if it arises from a
bilinear equation as above).

\subsubsection{The 3SS-condition}
\begin{defn}  (Hirota integrability)

Assume that the bilinear equation \eqref{eq:generic} has 1SS's of the form
\[
F=1+\epsilon e^{\eta_j},\quad  \eta_j=\vec x\cdot\vec p_j+\eta_j^0,
\]
where the parameters $\vec p_j$ are only restricted by the DR $P(\vec
p_j)=0$. The equation is said to be {\em Hirota integrable} if it has
NSS's of the form
\[
F=1+\epsilon \sum_{j=1}^N e^{\eta_j}+\text{(finite number of
  h.o. terms in $\epsilon$)}
\]
{without any further conditions on the parameters $\vec p_j$ of the
individual solitons.}
\end{defn}
The assumption that there are no further restrictions is important,
because any equation has multi-soliton solutions for some restricted set
of parameters.

If we now apply this principle to the 3SS we start
with the ansatz
\begin{eqnarray}
  F&=&1+\epsilon\, e^{\eta_1}+\epsilon\, e^{\eta_2}+\epsilon\,
  e^{\eta_3}\nonumber\\
  &&+\epsilon^2\, A_{12}e^{\eta_1+\eta_2}+\epsilon^2\,
  A_{23}e^{\eta_2+\eta_3}+\epsilon^2\, A_{31}e^{\eta_3+\eta_1}
+\epsilon^3\, A_{12}A_{23}A_{13}e^{\eta_1+\eta_2+\eta_3},\label{eq:3SS}
\end{eqnarray}
which is fixed by the requirement that if in a NSS any soliton goes
far away, the rest should look like the (N-1)SS. (In practice ``going
away'' means either $e^{\eta_k}\to 0$ or $e^{\eta_k}\to \infty$.)
This means that there is no freedom left: parameters are restricted only
by the dispersion relation \eqref{eq:genDR} and the phase factors
$A_{ij}$ were given already in \eqref{eq:A12}.
Thus the existence of three-soliton solutions is not automatic, but
rather it imposes severe requirements on the polynomial $P$.

The above definition of integrability can be used as {\em a method for
  searching for new integrable equations}. In practice it means that
one should find polynomials $P$ that verify the
``three-soliton-condition''(3SC), which follows when the ansatz
\eqref{eq:3SS} is substituted into \eqref{eq:generic}, namely
\begin{eqnarray}
&&\sum_{\sigma_i=\pm1} P(\sigma_1\vec p_1+\sigma_2\vec
p_2+\sigma_3\vec p_3) P(\sigma_1\vec p_1-\sigma_2\vec p_2)\nonumber
\\ \label{eq:P3SC} &&\qquad\hspace{2cm}\times
P(\sigma_2\vec p_2-\sigma_3\vec p_3) P(\sigma_1\vec p_1-\sigma_3\vec
p_3)=0,
\end{eqnarray}
on the manifold $P(\vec p_i)=0,\forall i$. {\em If} the 3SC is
satisfied then the equation most likely has NSS of the form
\begin{equation}
  \label{eq:cNSS}
  F=\sum_{\mu_i\in\{0,1\}}\exp\left[\sum_{i>j}^{(N)}a_{ij}\,\mu_i\,\mu_j+
    \sum_{i=1}^N\mu_i\,\eta_i\right],\quad\text{where}\quad \exp(a_{ij})=A_{ij}.
\end{equation}

\subsubsection{Continuous searches}
Condition \eqref{eq:P3SC} can be easily converted for symbolic
computation, for example in REDUCE \cite{RED} one can use the {\tt
  LET}-rule for the dispersion relations (which in REDUCE is evaluated
to the end). In 1987-1988 one of the present authors (JH) used this
method to search for new integrable equations of KdV, mKdV, sG and nlS
type \cite{search,search_nls}, for which conditions similar to
\eqref{eq:P3SC} can be derived. Among the (very few) new equations
found in this way were
\begin{align}
  (D_x^3D_t+{ aD_x^2}+D_tD_y)F\cdot F &=0,\label{E:HSI}  \\
  (D_x^4-D_xD_t^3+aD_x^2+bD_xD_t+cD_t^2)F\cdot F
  &=0,\label{E:my}
\end{align}
and
\begin{equation}
\left\{\begin{array}{rcl}
(i\alpha D_x^3+3D_xD_y-2iD_t+c)\, G\cdot F&=&0,\\
\bigl[a(\alpha^2D_x^4-3D_y^2+4\alpha D_xD_t)+bD_x^2\bigr]\, F\cdot F&=&|G|^2.
\end{array}\right.
\label{E:kpds}
\end{equation}
The pair of equations \eqref{E:kpds} is of the nonlinear
Shr\"odinger type and in fact combines the two most important
$(2+1)$-dimensional equations, Davey-Stewartson and
Kadomtsev-Petviashvili equations, but only their DSII and KPI
variants.

\subsection{Hirota's bilinear formalism for lattice equations}
We now turn to discrete lattice equations and as a model we use the
continuous situation described above. The various differences between
continuous and discrete models (such as the lack of Leibniz rule in
the latter) make things now much more difficult. We will only consider
P$\Delta$E's defined on the 2-dimensional cartesian lattice.

\subsubsection{Discrete Hirota bilinear form}
Before we can proceed to equations we must have a definition of the
discrete Hirota bilinear form. The guiding property is still the
gauge-invariance \eqref{eq:gauge}, and thus assuming we have lattice
functions $f_j(n,m)$, we require gauge-invariance under
\begin{equation}  \label{eq:dgauge}
f_j(n,m)\mapsto  f'_j(n,m)=A^nB^m\, f_j(n,m).
\end{equation}

Guided by the requirement of invariance under
\eqref{eq:dgauge} we have
\begin{defn}  (Discrete Hirota bilinear form)
If the equation can be written as
  \begin{equation*}
    \label{eq:HB}
    \sum_j\, c_j\, f_{j}(n+\nu_{j}^+,m+\mu_{j}^+)\,
g_{j}(n+\nu_{j}^-,m+\mu_{j}^-)=0
  \end{equation*}
  where the index sums $\mu_{j}^++\mu_{j}^-=\mu^s$,
  $\nu_{j}^++\nu_{j}^-=\nu^s$ do not depend on $j$,
we say it has {\em discrete Hirota bilinear form}.
\end{defn}

Note that
\begin{equation*}
e^{aD_x}\,f(x)\cdot g(x)= \left.
  e^{a(\partial_x-\partial_{x'})}\,f(x)g(x')\right|_{x'=x} = f(x+a)g(x-a)\ .
\end{equation*}
Therefore the discrete version of $P(D)f\cdot f=0$ should be an even
function built up from exponentials of the Hirota derivative.  For
example a sum containing terms like $\cosh(D_x)$ or
$\sinh(D_x)\sinh(D_t)$ etc., is acceptable.

\subsubsection{The Hirota and Miwa equations}
The first results on discrete bilinear soliton equations were obtained
by Hirota in a series of papers in 1977 \cite{Hir77}. Hirota's aim was
to discretize known soliton equations so that their multi-soliton
solutions would still have more or less the same form. In this way he
was able to derive integrable 2D lattice equations for KdV, Toda, and
sine-Gordon equations. The culmination of this series was the
``Discrete Analogue of a Generalized Toda Equation'' (DAGTE) or
``Hirota-equation'' \cite{Hir81}, which in the fully 3D version
(discrete bilinear KP equation) can be written as \cite{Miw82}
\begin{subequations}
 \begin{equation}
  \label{eq:dcDAGTE}
  a(b-c)\,\tau_{n+1,m,k}\tau_{n,m+1,k+1}+  b(c-a)\,\tau_{n,m+1,k}\tau_{n+1,m,k+1}+
 c(a-b)\,\tau_{n,m,k+1}\tau_{n+1,m+1,k}=0.
\end{equation}
This equation has soliton solutions of the form \eqref{eq:cNSS}, but with
discretized PWF's given by \cite{Miw82,DJM83}
\begin{equation}
  \label{eq:DAGTE-NSS}
  e^{\eta_i}=\left(\frac{1-aq_i}{1-ap_i}\right)^n
\left(\frac{1-bq_i}{1-bp_i}\right)^m
\left(\frac{1-cq_i}{1-cp_i}\right)^k,\quad
A_{ij}=\frac{(p_i-p_j)(q_i-q_j)}{(p_i-q_j)(q_i-p_j)}.
\end{equation}
\end{subequations}

In the early 1980's the continuum bilinear approach was developed to a
complete theory by Kyoto school, and it was very soon applied to the
discrete case as well \cite{Miw82,DJM83}. We note in particular the
generalization of Hirota's equation \eqref{eq:dcDAGTE} by Miwa (the
discrete BKP equation)
\cite{Miw82}
\begin{subequations}
\begin{align}
  &(a+b)(a+c)(b-c)\,\tau_{n+1,m,k}\tau_{n,m+1,k+1}
+ (b+c)(b+a)(c-a)\,\tau_{n,m+1,k}\tau_{n+1,m,k+1}\nonumber\\
+ &(c+a)(c+b)(a-b)\,\tau_{n,m,k+1}\tau_{n+1,m+1,k}
+ (a-b)(b-c)(c-a)\,\tau_{n+1,m+1,k+1}\tau_{n,m,k}=0.   \label{eq:HM}
\end{align}
Also this equation has soliton solutions of the form \eqref{eq:cNSS},
except that now \cite{Miw82,DJM83}
  \label{eq:Miwa-NSS}
\begin{eqnarray}
  e^{\eta_i}&=&
\left(\frac{(1-ap_i)(1-aq_i)}{(1+ap_i)(1+aq_i)}\right)^n
\left(\frac{(1-bp_i)(1-bq_i)}{(1+bp_i)(1+bq_i)}\right)^m
\left(\frac{(1-cp_i)(1-cq_i)}{(1+cp_i)(1+cq_i)}\right)^k,\label{eq:Miwa-NSS1}\\
A_{ij}&=&\frac
{(p_i-p_j)(p_i-q_j)(q_i-p_j)(q_i-q_j)}
{(p_i+p_j)(p_i+q_j)(q_i+p_j)(q_i+q_j)}.\label{eq:Miwa-NSS2}
\end{eqnarray}
\end{subequations}

\subsubsection{Application of Hirota's method to some P$\Delta$E's}
Hirota's direct method has also been applied to construction of
soliton solutions to nonlinear lattice equations. There the first
problem is to construct the dependent variable transform that takes
one from the nonlinear form to the Hirota bilinear form with gauge
invariance. This approach was used for most of the equations in the
ABS list in \cite{AHN08,HZ09}. For example, in the case of the H1
equation
\begin{equation}
{\rm H}1 \equiv (u-\widehat{\widetilde{u}})(\t{u}-\h{u})+(a^2-b^2)=0,
\label{H1}
\end{equation}
one is led (after studying the 1SS and 2SS) to the dependent variable
transformation
\begin{equation}
  u_{n,m}^{(NSS)}  = a n + b m +\lambda
  - \frac{g_{n,m}}{f_{n,m}},
\label{depvar-H1}
\end{equation}
which yields the bilinearization
\begin{align*}
\mathrm{H}1=-\bigl[\mathcal{H}_1+(a-b)f\widehat{\widetilde{ f}}\bigr]
\bigl[\mathcal{H}_2+(a+b)\h f\t f\bigr]/(f\h f\t f \widehat{\widetilde{ f}})
+(a^2-b^2),
\end{align*}
where
\begin{subequations}  \label{eq:bil-H1}
\begin{eqnarray}
  \mathcal{H}_1 &\equiv & \h g\t f-\t g\h f+(a-b)(\h f\t f-
  f\widehat{\widetilde{ f}})=0,
  \label{eq:bil-H1-1}\\
  \mathcal{H}_2 &\equiv & g\widehat{\widetilde{
      f}}-\widehat{\widetilde{ g}} f+(a+b)(f\widehat{\widetilde{
      f}}-\h f\t f)=0
\label{eq:bil-H1-2}.
\end{eqnarray}
\end{subequations}

\section{Searching for integrable bilinear lattice equations}
Following the continuous example we would like to search for
integrable lattice equations by the three-soliton condition. Since we
will here only consider 1-component equations, we can still take
\eqref{eq:generic} as the basic class of equations, but now $P$ would
have to be a sum of exponentials.

\subsection{Organizational questions}
The first question is: How to organize such a search? In the continuous
case we used the total degree in the bilinear derivative and then the
degrees in various coordinates. The problem in the discrete case is
that there are so many ways to discretize a derivative and thus there
will be many more cases.

\begin{figure}
\begin{center}
\setlength{\unitlength}{0.0034in}
\begin{picture}(1000,600)(0,50)
\drawline(0,100)(1200,100)
\drawline(0,200)(1200,200)
\drawline(0,300)(1200,300)
\drawline(0,400)(1200,400)
\drawline(0,500)(1200,500)
\drawline(0,600)(1200,600)
\drawline(50,650)(50,50)
\drawline(150,650)(150,50)
\drawline(250,650)(250,50)
\drawline(350,650)(350,50)
\drawline(450,650)(450,50)
\drawline(550,650)(550,50)
\drawline(650,650)(650,50)
\drawline(750,650)(750,50)
\drawline(850,650)(850,50)
\drawline(950,650)(950,50)
\drawline(1050,650)(1050,50)
\drawline(1150,650)(1150,50)

\thicklines

\drawline(150,600)(250,500)
\drawline(150,500)(250,600)
\put(150,600){\circle*{15}}
\put(150,500){\circle*{15}}
\put(250,600){\circle*{15}}
\put(250,500){\circle*{15}}
\put(80,525){\makebox(0,0)[lb]{\small$a)$}}

\drawline(150,400)(350,200)
\drawline(150,200)(350,400)
\drawline(150,300)(350,300)
\drawline(250,400)(250,200)
\put(150,400){\circle*{15}}
\put(150,300){\circle*{15}}
\put(150,200){\circle*{15}}
\put(250,400){\circle*{15}}
\put(250,300){\circle*{15}}
\put(250,200){\circle*{15}}
\put(350,400){\circle*{15}}
\put(350,300){\circle*{15}}
\put(350,200){\circle*{15}}
\put(80,325){\makebox(0,0)[lb]{\small$b)$}}

\drawline(550,500)(850,200)
\drawline(550,200)(850,500)
\drawline(550,300)(850,400)
\drawline(550,400)(850,300)
\drawline(650,500)(750,200)
\drawline(650,200)(750,500)
\put(550,500){\circle*{15}}
\put(550,400){\circle*{15}}
\put(550,300){\circle*{15}}
\put(550,200){\circle*{15}}
\put(650,500){\circle*{15}}
\put(650,400){\circle*{15}}
\put(650,300){\circle*{15}}
\put(650,200){\circle*{15}}
\put(750,500){\circle*{15}}
\put(750,400){\circle*{15}}
\put(750,300){\circle*{15}}
\put(750,200){\circle*{15}}
\put(850,500){\circle*{15}}
\put(850,400){\circle*{15}}
\put(850,300){\circle*{15}}
\put(850,200){\circle*{15}}
\put(480,325){\makebox(0,0)[lb]{\small$c)$}}

\put(1050,500){\circle*{15}}
\put(1050,400){\circle*{15}}
\put(1050,300){\circle*{15}}
\put(1050,200){\circle*{15}}
\put(1150,500){\circle*{15}}
\put(1150,400){\circle*{15}}
\put(1150,300){\circle*{15}}
\put(1150,200){\circle*{15}}
\drawline(1050,500)(1150,200)
\drawline(1050,400)(1150,300)
\drawline(1050,300)(1150,400)
\drawline(1050,200)(1150,500)
\put(980,325){\makebox(0,0)[lb]{\small$d)$}}
\end{picture}
\end{center}
\caption{Possible stencils for 2D bilinear equations. Case a) only
  contains 1 equation \eqref{eq:small}, Case b) is analyzed in this
  paper, Case d) and similar models are analyzed in
  \cite{RIAM}.\label{F:rad}}
\end{figure}
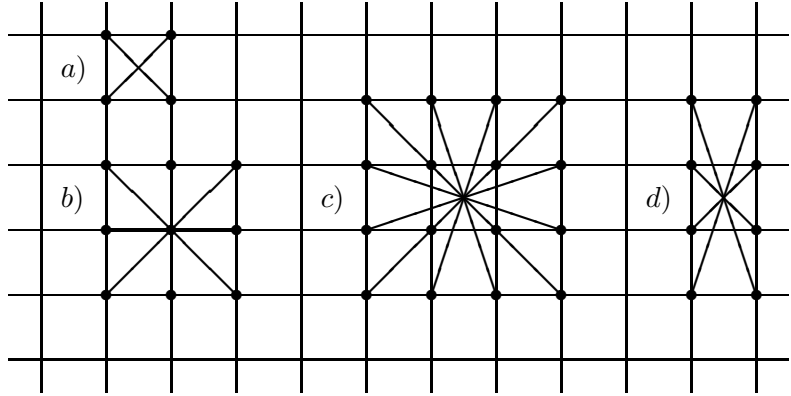

One way to proceed is by the radius of
points needed, see Figure \ref{F:rad}.
For sub-figure \ref{F:rad}a) the corresponding bilinear equation will be
\begin{equation}
  \label{eq:small}
  c_1\,f_{n,m}f_{n+1,m+1}+ c_2\,f_{n+1,m}f_{n,m+1}=0,
\end{equation}
since these are the only combinations of
$f_{n+\nu,m+\mu}f_{n+\nu',m+\mu'}$ with non-negative
$\mu,\mu',\nu,\nu'$ and sum $\mu+\mu'=1,\nu+\nu'=1$. Since we want
$f\equiv1$ to be a solution we must have $c_1+c_2=0$ and then there are no
parameters, except the trivial scaling.

In this paper we will consider the arrangement given in sub-figure
\ref{F:rad}b) which corresponds to the difference equation
\begin{align}
g_8&\, f_{n+1,m+1}\,f_{n-1,m-1}+g_6\,
f_{n+1,m-1}\,f_{n-1,m+1}\nonumber \\
+&g_7\, f_{n+1,m}\,f_{n-1,m}+h_6
\,f_{n,m+1}\,f_{n,m-1}+h_7\,f_{n,m}^2=0. \label{eq:genr2eq}
\end{align}
In terms of Hirota's $D$-derivatives this can be written as
\eqref{eq:generic}, with the function
\[
P(X,Y):=g_8\,e^{X+Y}+g_6\,e^{X-Y}+g_7\,e^{X}+
h_6\,e^{Y}+h_7.
\]
However, since the $D$-operator is antisymmetric and $P$ operates on a
symmetric target $f\cdot f$, only the symmetric part of $P$ is relevant. Thus
the proper form is
\begin{equation}
  \label{eq:P9}
P(X,Y):=g_8\,\cosh(X+Y)+g_6\,\cosh(X-Y)+g_7\,\cosh(X)+
h_6\,\cosh(Y)+h_7.
\end{equation}

Since equation \eqref{eq:tail} still holds, the construction of 1SS
and 2SS goes as usual, resulting with formulae
(\ref{eq:1SS}-\ref{eq:A12}), with $\eta_j=p_jn+q_jm+\eta_j^0$. In
particular, from the 0SS $f_{n,m}=1,\,\forall n,m$ we get the condition
\begin{equation}
g_6+g_7+g_8+h_6+h_7=0,\label{eq:c0SS}
\end{equation}
and from the 1SS $f_{n,m}=1+e^{\eta_j}$ we get the DR
$P(p_j,q_j)=0$, i.e.,
\begin{equation}
  \label{eq:DRcosh}
g_8\,\cosh(p_j+q_j)+g_6\,\cosh(p_j-q_j)+g_7\,\cosh(p_j)+
h_6\,\cosh(q_j)+h_7=0,
\end{equation}
while the formulae for $A_{ij}$ \eqref{eq:A12} and for the 3SC
\eqref{eq:P3SC} hold as written.

The practical problem is the following: How to implement the
dispersion relation \eqref{eq:DRcosh} while keeping the phase factor
and the 3SC manageable?  It seems that in order to implement the DR we
must rationalize it somehow, but then we will loose the additive form
of the phase factor and the 3SC.

As the first step we replace
\[
e^{p_j}=p_{j}',e^{q_j}=q_j'
\]
and introduce the function
\[
P'(x,y):=(xy)^{-1}[g_8\, (x^2 y^2 + 1) + g_6\, (x^2 + y^2) + g_7\, y (x^2 +
  1) + h_6\, x(y^2 + 1) +   2 h_7\, x y],
\]
after which the dispersion relation is still $P'(p'_j,q'_j)=0$, but
\[
A_{ij}=-\frac{P'(p'_i/p'_j,q'_i/q'_j)}{P'(p'_ip'_j,q'_iq'_j)},
\]
and correspondingly in the 3SC. This does not look bad, except that
the DR is still hard to implement. Next we continue with
\begin{equation}
  \label{eq:tra1}
p_j'=\tfrac{P_j-1}{P_j+1},\quad q_j'=\tfrac{Q_j-1}{Q_j+1}
\end{equation}
so that
\begin{equation}
  \label{eq:1SSd2}
  f_{n,m}=1+ c_j \left(\tfrac{P_j-1}{P_j+1}\right)^n
\left(\tfrac{Q_j-1}{Q_j+1}\right)^m.
\end{equation}
Now the DR gets a nice form
\begin{equation}
  \label{eq:DRd2}
  {\mathcal P}(P_j,Q_j):=(P_j^2-1)^{-1}(Q_j^2-1)^{-1}
  (o_1\, P_j^2+2o_3\, P_jQ_j+o_2\, Q_j^2  +o_4)  =0
\end{equation}
where
\[
\begin{array}{l}
o_1=g_7+h_7,\\ o_2=h_6+h_7,\\ o_3=g_6-g_8,\\ o_4=g_7+h_6,
\end{array}\quad\text{with inverse}\quad
\begin{array}{l}
g_6=\tfrac14( - o_1 - o_2 + 2o_3 - o_4),\\
g_7=\tfrac12(o_1 - o_2 + o_4),\\
g_8=\tfrac14( - o_1 - o_2 - 2o_3 - o_4),\\
h_6=\tfrac12( - o_1 + o_2 + o_4),\\
h_7=\tfrac12(o_1 + o_2 - o_4).
\end{array}
\]
However, the simple additive property in $A_{ij}$ is lost, we find instead
\begin{equation}\label{eq:A12d2}
A_{ij}=-\frac{{\mathcal P}
\left(\tfrac{1-P_iP_j}{P_i-P_j},\tfrac{1-Q_iQ_j}{Q_i-Q_j}\right)}
{{\mathcal P}\left(\tfrac{1+P_iP_j}{P_i+P_j},\tfrac{1+Q_iQ_j}{Q_i+Q_j}\right)}
\end{equation}
and  the 3SC is still much more complicated.

In the following we use the parametrization \eqref{eq:1SSd2}, which
implies
(\ref{eq:DRd2},\ref{eq:A12d2}).

\subsection{Results}
The search was performed by computing the 3SS subject to the dispersion
relation
\begin{equation}
  \label{eq:DR}
o_1\, P_j^2+2o_3\, P_jQ_j+o_2\, Q_j^2 +o_4=0.
\end{equation}
The dispersion relation was implemented using {\tt LET}-rule of REDUCE
\cite{RED} on the leading monomial (in ``gradlex'' ordering). In REDUCE
such rules are evaluated to the end.

\begin{itemize}
\item For the generic case we may assume that $o_1\neq0$ and scale it
  to $o_1=1$, then we can use the rule: {\tt for all n let
    pp(n)\^{}2=-2*o3*pp(n)*qq(n)-o2*qq(n)\^{}2-o4;} The resulting 3SC
  is rather long but the coefficients of various powers of $Q_i$ and
  first powers of $P_i$ are relatively simple and one can proceed by
  solving them one by one.

\item If $o_1=0$ then we may assume $o_3\neq 0$ and scale it to $o_3=-1/2$.
Then we can solve for $P_j$ and use the substitution {\tt for all n
  let pp(n)=(o2*qq(n)\^{}2+o4)/qq(n);} Some of these cases are
$n\leftrightarrow m$ reflections of $o_1=1,o_2=0$.
\end{itemize}

The process branches somewhat but when the results
are collected and sub-cases eliminated we get the following cases:

{\small
\begin{enumerate}

\item $o_1=1,o_2=2o_3 - o_4 - 1$ %80
\[
  f_{n+1,m+1} f_{n-1,m-1} o_3 + f_{n+1,m} f_{n-1,m} (o_3 - o_4 -
  1) + f_{n,m+1} f_{n,m-1} ( - o_3 + 1) + f_{n,m}^2 ( - o_3 +
  o_4)=0.
\]

\item $ o_1=1,o_2= - 2 o_3 - o_4 - 1 $ %85
\[
f_{n+1,m} f_{n-1,m} (o_3 + o_4 + 1) +
  f_{n+1,m-1} f_{n-1,m+1} o_3 - f_{n,m+1} f_{n,m-1} (o_3 + 1) -
  f_{n,m}^2 (o_3 + o_4)=0.
\]
\item $ o_1=1,o_2=1,o_4=0 $
\[
f_{n+1,m+1} f_{n-1,m-1} (o_3 + 1) +
  f_{n+1,m-1} f_{n-1,m+1} ( - o_3 + 1) - 2 f_{n,m}^2=0.
\]
\item $ o_1=1,o_2=0,o_4=1$
\[
f_{n+1,m+1} f_{n-1,m-1} (o_3 + 1) - 2 f_{n+1,m} f_{n-1,m} + f_{n+1,m-1}
f_{n-1,m+1} ( - o_3 + 1)=0.
\]
\item $ o_1=0,o_3=-1/2,o_4=o_2$
\[
f_{n+1,m+1} f_{n-1,m-1} ( - 2 o_2 + 1) + f_{n+1,m-1} f_{n-1,m+1} ( - 2 o_2 - 1
) + 4 f_{n,m+1} f_{n,m-1} o_2=0.
\]
\item $ o_1=0,o_3=-1/2,o_4=1-o_2$
\[
f_{n+1,m} f_{n-1,m} ( - 2 o_2 + 1) - f_{n+1,m-1} f_{n-1,m+1} + f_{n,m+1}
 f_{n,m-1} + f_{n,m}^2 (2 o_2 - 1)=0.
\]
\item $ o_1=0,o_3=-1/2,o_4=-1-o_2$
\[
f_{n+1,m+1} f_{n-1,m-1} + f_{n+1,m} f_{n-1,m} ( - 2 o_2 - 1) - f_{n,m+1}
 f_{n,m-1} + f_{n,m}^2 (2 o_2 + 1)=0.
\]
\end{enumerate}
}

For further analysis we restore the freedom in the scaled parameters
and combine cases to the extent possible. The final results can then
be divided into the following cases:
\renewcommand{\dashlinestretch}{200}

\begin{figure}[t]
\setlength{\unitlength}{0.003in}
\begin{picture}(300,600)(50, 0)
\thinlines
\drawline(100,100)(600,100)
\drawline(100,200)(600,200)
\drawline(100,300)(600,300)
\drawline(100,400)(600,400)
\drawline(100,500)(600,500)
\drawline(150,550)(150,50)
\drawline(250,550)(250,50)
\drawline(350,550)(350,50)
\drawline(450,550)(450,50)
\drawline(550,550)(550,50)
\thicklines
\dashline{10}(250,400)(450,200)
\dashline{10}(355,200)(355,400)
\dashline{10}(250,305)(450,305)
\put(450,300){\circle*{18}}
\put(450,200){\circle*{18}}
\put(350,400){\circle*{18}}
\put(350,300){\circle*{25}}
\put(250,300){\circle*{18}}
\put(350,200){\circle*{18}}
\put(250,400){\circle*{18}}
\put(300,000){\makebox(0,0)[cc]{{\bf Case 1b)}}}
\end{picture}
\begin{picture}(300,600)(-250, 0)
\thinlines
\drawline(100,100)(600,100)
\drawline(100,200)(600,200)
\drawline(100,300)(600,300)
\drawline(100,400)(600,400)
\drawline(100,500)(600,500)
\drawline(150,550)(150,50)
\drawline(250,550)(250,50)
\drawline(350,550)(350,50)
\drawline(450,550)(450,50)
\drawline(550,550)(550,50)
\thicklines
\dashline{10}(250,200)(450,400)
\dashline{10}(250,400)(450,200)
\put(450,200){\circle*{18}}
\put(450,400){\circle*{18}}
\put(250,400){\circle*{18}}
\put(350,300){\circle*{25}}
\put(250,200){\circle*{18}}
\put(300,000){\makebox(0,0)[cc]{{\bf Case 2)}}}
\end{picture}
\begin{picture}(300,600)(-550, 0)
\thinlines
\drawline(100,100)(600,100)
\drawline(100,200)(600,200)
\drawline(100,300)(600,300)
\drawline(100,400)(600,400)
\drawline(100,500)(600,500)
\drawline(150,550)(150,50)
\drawline(250,550)(250,50)
\drawline(350,550)(350,50)
\drawline(450,550)(450,50)
\drawline(550,550)(550,50)
\thicklines
\dashline{10}(250,200)(450,400)
\dashline{10}(250,305)(450,305)
\dashline{10}(250,400)(450,200)
\put(450,200){\circle*{18}}
\put(450,400){\circle*{18}}
\put(250,300){\circle*{18}}
\put(450,300){\circle*{18}}
\put(250,400){\circle*{18}}
\put(250,200){\circle*{18}}
\put(300,000){\makebox(0,0)[cc]{{\bf Case 3a)}}}
\end{picture}
\caption{Points involved in the equations with 3SS. Case 1a is
  obtained by a $90^o$ rotation from Case 1b, and Case 3b the same way
  from Case 3a.}
\label{Fig1}
\end{figure}
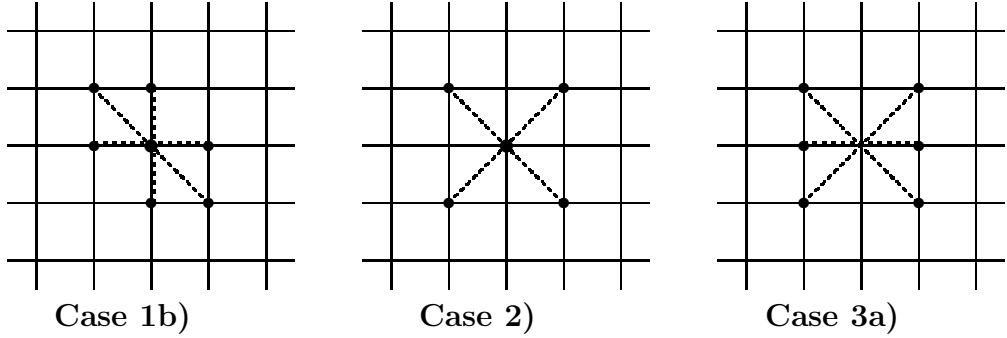

\paragraph{Case 1:} $o_1+o_2+o_4=\pm 2o_3$: %g1a,g1b or 80,85
\begin{subequations}
\begin{equation}
  \label{eq:c1a}
f_{n+1,m+1} f_{n-1,m-1} o_3 + f_{n+1,m} f_{n-1,m} (o_2 - o_3) +
f_{n,m+1} f_{n,m-1} (o_1 - o_3) + f_{n,m}^2 (o_3 -o_1- o_2)=0.
\end{equation}
\begin{equation}
  \label{eq:c1b}
f_{n+1,m-1} f_{n-1,m+1} o_3 - f_{n+1,m} f_{n-1,m} (o_2 + o_3) -
f_{n,m+1} f_{n,m-1} (o_1 + o_3) + f_{n,m}^2 (o_3 + o_1+o_2)=0.
\end{equation}
\end{subequations}
These are related by $m \leftrightarrow -m$; due to \eqref{eq:1SSd2}
this corresponds to $Q \leftrightarrow -Q$, or in the dispersion
relation,  $o_3 \leftrightarrow -o_3$. The associated diagram is
given in Figure \ref{Fig1}.

For $o_1=1$ this contains items 1 and 2 of the previous list, and
for $o_1=0,o_3=-1/2,o_4=-1-o_2$ the first one also contains item 7 and
for $o_1=0,o_3=-1/2,o_4=1-o_2$ the second one contains item 6.

\paragraph{Case 2:} $o_1=o_2\neq 0,o_4=0$ (item 3 from the previous
list): %g2a or 64
\begin{equation}
  \label{eq:c2}
f_{n+1,m+1} f_{n-1,m-1} (o_3 + o_1) +
  f_{n+1,m-1} f_{n-1,m+1} ( - o_3 + o_1) - 2 o_1 f_{n,m}^2=0.
\end{equation}
This is the Toda-lattice (center + cross, see Figure \ref{Fig1}).

\paragraph{Case 3:} $o_2=0,o_4=o_1$ %g3a, or 42 and x24
or $o_1=0,o_4=o_2$ (or: $o_1+o_2=o_4,\,o_1-o_2=\pm o_4$).
%\[f_{n+1,m+1} f_{n-1,m-1} (o_3 + o_1) + f_{n+1,m-1} f_{n-1,m+1} (o_1 -
%o_3) - 2o_1 f_{n+1,m} f_{n-1,m}=0.\]
\begin{subequations}
\begin{equation}
  \label{eq:c3a}
f_{n+1,m+1} f_{n-1,m-1} (o_3 + o_1) + f_{n+1,m-1} f_{n-1,m+1} (o_1 -
o_3) - 2o_1 f_{n+1,m} f_{n-1,m}=0.
\end{equation}
\begin{equation}
  \label{eq:c3b}
f_{n+1,m+1} f_{n-1,m-1} (o_2 + o_3) + f_{n+1,m-1} f_{n-1,m+1} (o_2 - o_3
) - 2 o_2 f_{n,m+1} f_{n,m-1}=0.
\end{equation}
\end{subequations}
These contain item 4 and its $n\leftrightarrow m$ reflection, item 5.
Both are three term equations with two diagonals and a vertical or a
 horizontal, see Figure \ref{Fig1}.  They correspond to Hirota's
discretization of the KdV equation \cite{Hir77}.

\section{Post processing}
In this section we will take a closer look on the equations obtained,
and in particular on how the soliton solutions look and how the final
results can be obtained as a reduction from the Hirota or Miwa equations.
In the following the parameters $o_i$ are from the search, $a,b,c$ are
those used in the Hirota and Miwa equations \eqref{eq:dcDAGTE},
\eqref{eq:HM}, and $\alpha,\beta$ are the canonical parameters in
terms of which the PWF's have the simplest form.

\subsection{Case 2:}
\subsubsection{A better parameterization}
The dispersion relation for Case 2 reads
\begin{equation}
  \label{eq:DRc2}
o_1\, P^2+2o_3\, PQ+o_1\, Q^2=0.
\end{equation}
Parameterizing $o_1(\neq 0),\,o_3$ as
\begin{equation}
  \label{eq:ac2albe}
o_1=2\alpha\beta,\quad o_3=-(\alpha^2+\beta^2),
\end{equation}
the DR factorizes as
\begin{equation}
  \label{eq:DRc2f}
(\alpha P-\beta Q)(\beta P-\alpha Q)=0.
\end{equation}
Let us fix $\alpha,\,\beta$ as determined from \eqref{eq:ac2albe}. In
the generic case $\alpha^2\neq\beta^2$ and we have in fact two different kinds
of solitons, depending on which factor of \eqref{eq:DRc2f} is
satisfied. Thus we may choose either
\[
Q_j=\beta/k_j,\quad P_j=\alpha/k_j,
\]
or
\[
Q_j=\alpha/k_j,\quad P_j=\beta/k_j,
\]
and obtain the PWF
\begin{equation}
  \label{eq:res-c2x}
e^{\eta_i}=\left(\frac{\alpha_i-k_i}{\alpha_i+k_i}\right)^n
\left(\frac{\beta_i-k_i}{\beta_i+k_i}\right)^m,
\text{ where $(\alpha_i,\beta_i)=(\alpha,\beta)$ or $(\beta,\alpha)$. }
\end{equation}
The phase factor $A_{ij}$ has the canonical form
\begin{equation}
  \label{eq:A12-can}
A_{ij}=\left(\frac{k_i-k_j}{k_i+k_j}\right)^2.
\end{equation}

\subsubsection{Reduction from Hirota's equation}
Equation \eqref{eq:c2} is obtained from \eqref{eq:dcDAGTE} by the
reduction
\begin{equation}
  \label{eq:redudef2}
\tau(n+1,m+1,k)=\tau(n,m,k+1),
\end{equation}
which we use to define a new function $f$ by:
\[
\tau_{n,m,k}= f_{n+m+2k,n-m} .
\]
Thus we define the new independent variables by
\[
n':=m+n+2k-2,\quad m':=n-m
\]
and then it is easy to see that the 6 corner points of the cube on
which \eqref{eq:dcDAGTE} is defined, namely
$\{(100),(011),(010),(101),(001),(110)\}$ will project to
$\{(-1,1),(1,-1),(-1,-1),(1,1),\linebreak (0,0),(0,0)\}$, displayed at
the center of Figure \ref{Fig1}.  With this definition one obtains
\eqref{eq:c2} for $f_{n'm'}$ with,
\[
o_1=\tfrac12c(b-a),\quad o_3=\tfrac12(c(a+b)-2ab).
\]

\subsubsection{Reduction of solution\label{S413}}
For the NSS of DAGTE given in \eqref{eq:DAGTE-NSS} the reduction
\eqref{eq:redudef2} implies
\[
\left(\frac{1-aq_i}{1-ap_i}\right)
\left(\frac{1-bq_i}{1-bp_i}\right)=
\left(\frac{1-cq_i}{1-cp_i}\right),
\]
which has the nontrivial solution
\begin{equation}
  \label{eq:c1subq}
q_i=\frac{abp_i-a-b+c}{ab(c p_i-1)}.
\end{equation}
Here we must assume that $ab(a-c)(b-c)\neq 0$.
After this the PWF of \eqref{eq:DAGTE-NSS} gets the form
\begin{equation}
  \label{eq:pwf2alt}
e^{\eta_i}=
\left(\sqrt{\tfrac{a-c}b}\sqrt{\tfrac{b-c}a}\,
\frac{1}{cp_i-1}\right)^{n'+2}
\left(\sqrt{\tfrac{a-c}b}\sqrt{\tfrac{a}{b-c}}\,
\frac{bp_i-1}{ap_i-1}\right)^{m'}.
\end{equation}
We can write this in the canonical form for 2D systems
\begin{equation}
  \label{eq:res-c2}
e^{\eta_i}=\left(\frac{\alpha-k_i}{\alpha+k_i}\right)^n
\left(\frac{\beta-k_i}{\beta+k_i}\right)^m,\quad
\end{equation}
where instead of $p$ we have used $k$ defined by
\begin{equation}
  \label{eq:c2redefp}
p_i=\frac{1-uv}c\,\frac{k_i + s(u-v)(1+uv)}
{k_i + s(u-v)(1-uv)},\quad\text{where}\quad
u:=\sqrt{\tfrac{a-c}{b}},\quad v:=\sqrt{\tfrac{b-c}{a}},
\end{equation}
and then
\[
\alpha=s(uv-1)(u-v),\quad \beta=s(uv+1)(u+v).
\]
Here $s$ is an arbitrary parameter. Note that for this case we
have already derived PWF in the form \eqref{eq:res-c2x}, which allows
different kinds of solitons by $\alpha\leftrightarrow \beta$.  In the
above the sign change in $u$ corresponds to this freedom.

\subsubsection{Special cases}
The case $o_1=o_3$ corresponds to $a(b-c)=0$ and $o_1=-o_3$ to
$b(a-c)=0$, both of which break the  formulae in Sec.~\ref{S413}.
In fact in these cases the equation \eqref{eq:c2} only has two terms
and the system can be reduced to the 1D equation
$f_{N+1}f_{N-1}-f_N^2=0$.

The case $o_3=0$ corresponds to $\alpha=\pm i\beta$, which implies
$c=2ab/(a+b)$. The above formulae still hold for this special case.

\subsection{Case 3a:}
\subsubsection{A better parameterization}
For Case 3 the DR is ($o_1o_3\neq 0$)
\[
o_1P_i^2+2o_3P_iQ_i+o_1=0,\quad\text{i.e.,}\quad
Q_i=-\frac{o_1}{2o_3}(P_i+1/P_i),
\]
after which the PWF of \eqref{eq:1SSd2} becomes
\[
e^{\eta_i}=
\left(\frac{P_i-1}{P_i+1}\right)^n
\left(\frac{o_1P_i^2+2o_3P_i+o_1}{o_1P_i^2-2o_3P_i+o_1}\right)^{m},\quad
A_{ij}=\left(\frac{(P_iP_j-1)(P_i-P_j)}{(P_iP_j+1)(P_i+P_j)}\right)^2.
\]
In order to simplify this we define a new soliton parameter $k_i$ by
\[
k_i:=\frac{2\alpha P_i}{1+P_i^2},
\]
and furthermore let $o_3=-\alpha o_1/\beta$, after which we obtain
\begin{equation}
  \label{eq:res-c3a}
e^{\eta_i}=\left(\frac{\alpha-k_i}{\alpha+k_i}\right)^{n/2}
\left(\frac{\beta-k_i}{\beta+k_i}\right)^m,
\end{equation}
with $A_{ij}$ as in \eqref{eq:A12-can}.

The appearance of $n/2$ as the exponent in \eqref{eq:res-c3a} means
that the natural equation in this case is obtained by replacing
$n=2N$, i.e.,
\begin{equation}
  \label{eq:c3a-step1}
f_{N+1,m+1} f_{N,m-1} (o_3 + o_1) + f_{N+1,m-1} f_{N,m+1} (o_1 -
o_3) - 2o_1 f_{N+1,m} f_{N,m}=0,
\end{equation}
which involves points in the $2\times3$ sub-lattice. (In fact this
corresponds to the three-term equation (20) in \cite{RIAM}.)

\subsubsection{Reduction of equation}
Equation \eqref{eq:c3a} is obtained from Hirota's equation
\eqref{eq:dcDAGTE} by the reduction (see also \cite{DJM83}, Part III,
Example 1)
\begin{equation}
  \label{eq:redudef3a}
\tau(n,m+1,k+1)=\tau(n,m,k),
\end{equation}
which we use to define a new function $f$ by:
\[
\tau_{n,m,k}= f_{2n,m-k}.
\]
Defining further the new independent variables by
\[
n':=2n-1,\quad m':=m-k,
\]
one obtains \eqref{eq:c3a} for $f$ with,
\[
o_1=\tfrac12a(c-b),\quad o_3=\tfrac12(a(b+c)-2bc).
\]
In this case the 6 corner points used in \eqref{eq:dcDAGTE} are
projected to $\{(1,0),(-1,0),(-1,1),(1,-1),\linebreak (-1,-1),(1,1)\}$, as
displayed in rightmost part of Figure \ref{Fig1}.

\subsubsection{Reduction of the solution}
As for the PWF in \eqref{eq:DAGTE-NSS} the reduction
\eqref{eq:redudef3a} means
\[
\left(\frac{1-bq_i}{1-bp_i}\right)
\left(\frac{1-cq_i}{1-cp_i}\right)=1
\]
which has the nontrivial solution
\[
q_i=-p_i+\frac1b+\frac1c.
\]
After this the PWF of \eqref{eq:DAGTE-NSS} becomes
\[
e^{\eta_i}=
\left(\frac{a(b+c)-bc(ap_i+1)}{bc(ap_i-1)}\right)^{(n'+1)/2}
\left(\frac{b(1-cp_i)}{c(-1+bp_i)}\right)^{m'}.
\]
We can write this in the form \eqref{eq:res-c3a} after redefining
$p\to k$ by
\[
%p_i=\frac{k_i+1+\alpha}{c(\alpha+\beta+1)}\quad\text{ or }\quad
p_i=\tfrac1a k_i+\tfrac12[\tfrac1b+\tfrac1c]
\]
where we have used the parameterization
\[
a=c(\alpha+\beta+1),\quad b=c\frac{\alpha+\beta+1}{\alpha-\beta+1},
\quad
o_1=-\beta bc,\quad %-\beta c^2\frac{\alpha+\beta+1}{\alpha-\beta+1},\quad
o_3=\alpha bc.% c^2\frac{\alpha+\beta+1}{\alpha-\beta+1}.
\]
The phase factor $A_{ij}$ is as in \eqref{eq:A12-can}.

\subsubsection{Special cases}
The above construction works except if $o_1(o_1+o_3)(o_1-o_3)=0$,
corresponding to $abc(a-b)(a-c)(b-c)=0$.  In these cases equation
\eqref{eq:c3a}
reduces to a simple 2D equation, which can be normalized, e.g., to
$f_{N,M}f_{N+1,M+1}-f_{N+1,M}f_{N,M+1}=0$.

\subsection{Case 1b:}
We consider only Case 1b, i.e., $o_4=-(o_1+o_2+2o_3)$, Case 1a is
similar with some sign changes. Here we may assume that
$o_3(o_3+o_1)(o_3+o_2) \neq 0$ because otherwise the system reduces to
one of the 3-term equations discussed already.

\subsubsection{Details\label{S431}}
The dispersion relation is given by
\begin{equation}
  \label{eq:DRc1b}
o_1\, P_j^2+2o_3\, P_jQ_j+o_2\, Q_j^2=o_1+o_2+2o_3.
\end{equation}
In the generic case this can be resolved by
\begin{equation}
  \label{eq:par-c1-1}
  P_j=\tfrac1{1+\alpha}(k_j+\alpha/k_j),\quad
  Q_j=\tfrac1{1+\beta}(k_j+\beta/k_j),
\end{equation}
with
\begin{equation}
  \label{eq:par-c1-2}
  o_1=2\gamma\beta(1+\alpha)^2,\quad
  o_2=2\gamma\alpha(1+\beta)^2,\quad
  o_3=-\gamma(\alpha+\beta)(1+\alpha)(1+\beta),\quad
o_4=2\gamma(\alpha-\beta)^2.
\end{equation}
Here we must assume that $(1+\alpha)(1+\beta)\neq 0$. Furthemore, the
three special cases ($\alpha=\beta$), ($\alpha=0,\beta=1$) and
($\alpha=1,\beta=0$) may also be omitted as they lead to two-term
equations.

The PWF now becomes
\begin{subequations}\label{eq:solC1b}
\begin{equation}
  \label{eq:pwf-c1}
e^{\eta_i}=
  \left[\frac{(k_i-1)(k_i-\alpha)}{(k_i+1)(k_i+\alpha)}\right]^n
  \left[\frac{(k_i-1)(k_i-\beta)}{(k_i+1)(k_i+\beta)}\right]^m,
\end{equation}
and the phase factor
\begin{align}
  & A_{12}=\left(\frac{k_1-k_2}{k_1+k_2}\right)^2\times\label{eq:a12-c1}\\
  & \frac{k_1^2 k_2^2 (\alpha+\beta-1)-(k_1^2+k_2^2) \alpha \beta +
    k_1 k_2 (\alpha+\beta) (\alpha-1) (\beta-1)+\alpha \beta (-\alpha
    \beta+\alpha+\beta)} {k_1^2 k_2^2 (\alpha+\beta-1)-(k_1^2+k_2^2)
    \alpha \beta - k_1 k_2 (\alpha+\beta) (\alpha-1) (\beta-1)+\alpha
    \beta (-\alpha \beta+\alpha+\beta)}.\nonumber
\end{align}
\end{subequations}

Note that $\alpha,\beta,\gamma$ as determined from \eqref{eq:par-c1-2}
are not unique, but rather one can change $\alpha\mapsto1/\alpha,\,
\beta\mapsto1/\beta,\, \gamma\mapsto \gamma\alpha^2\beta^2$. If this
is accompanied with $k_i\mapsto 1/k_i$ then
equations (\ref{eq:pwf-c1},\ref{eq:a12-c1}) do not change.

\subsubsection{Reduction from Miwa's equation\label{S432}}
For Case 1 the equation has four terms and therefore one should be
able to obtain \eqref{eq:c1b} as a reduction of Miwa's equation
\eqref{eq:HM}.  Let us consider the reduction
\begin{subequations}
\begin{equation}
  \label{eq:redudef1b}
\tau(n,m,k)=\tau(n+1,m+1,k+1),
\end{equation}
and define
\begin{equation}\label{eq:redtf}
\tau_{n,m,k}=f_{n-k,k-m}
\end{equation}
and change variables by
\begin{equation}\label{eq:rednm}
n'=n-k,\,m'=k-m,
\end{equation}
\end{subequations}
This projection takes the corners of the cube on which \eqref{eq:HM}
is defined to the points given on the left of Figure \ref{Fig1}, in
particular the diagonal $\{(0,0,0),(1,1,1)\}$ projects to $(0,0)$.
Then for the equation we obtain \eqref{eq:c1b} for $n',m'$ (we drop
the primes in the following) with the
parameter correspondence
\begin{equation}
  \label{eq:par-c1-4}
o_1=2a(b^2-c^2),\,o_2=2b(c^2-a^2),\,
o_3=(a-b)(a+c)(b+c), \, o_4=-2c(a^2-b^2).
\end{equation}
Here we may assume that $(a^2-b^2)(b^2-c^2)(c^2-a^2)\neq 0$, because
otherwise we get a two-term equation \eqref{eq:small} with $c_1+c_2=0$.

%\subsubsection{Reduction of solution\label{S432}}
The solution to Miwa's equation \eqref{eq:HM} is given through 
(\ref{eq:Miwa-NSS1},\ref{eq:Miwa-NSS2}) and the reduction
\eqref{eq:redudef1b} implies the condition
\[
\frac{(1-ap_i)(1-aq_i)}{(1+ap_i)(1+aq_i)}
\frac{(1-bp_i)(1-bq_i)}{(1+bp_i)(1+bq_i)}
\frac{(1-cp_i)(1-cq_i)}{(1+cp_i)(1+cq_i)}=1
\]
and since $p_i+q_i\neq 0$ this leads to the bi-quadratic
\begin{equation}
  \label{eq:cond1b}
abc(ab+ac+bc)p_i^2q_i^2+abc(p_i^2+q_i^2)+(a+b)(b+c)(c+a)p_iq_i+a+b+c=0,
\end{equation}
which suggests an elliptic parameterization (if $abc\neq 0$). This
form of solution agrees with \eqref{eq:pwf-c1}, if in addition to
\eqref{eq:cond1b} we have
\begin{subequations}\label{eq:vert}
\begin{eqnarray}
  \frac{(k_i-1)(k_i-\alpha)}{(k_i+1)(k_i+\alpha)}
  =\frac{(1-ap_i)(1-aq_i)}{(1+ap_i)(1+aq_i)},\label{eq:vert1} \\
    \frac{(k_i-1)(k_i-\beta)}{(k_i+1)(k_i+\beta)}
    =\frac{(1+bp_i)(1+bq_i)}{(1-bp_i)(1-bq_i)}.\label{eq:vert2}
\end{eqnarray}
\end{subequations}
From these 3 equations one should, in principle, be able to determine
$p_i,\,q_i$ in terms $k_i$ and $\alpha,\beta$ in terms of $a,b,c$, but
this does not lead to simple expressions for them. In any case we have
verified that the equations are consistent provided that
$\alpha,\beta$ are determined from
\begin{equation}\label{rsalbe}
\hat r\alpha+\hat s\beta=1,\quad
\hat r/\alpha+\hat s/\beta=1,\quad
%r\frac1{\alpha}+s\frac1{\beta}=1,
\text{ where }\quad
\hat r:=\frac{b(a-c)^2}{c(a-b)^2},\quad
\hat s:=\frac{a(b-c)^2}{c(b-a)^2}.
\end{equation}
The same equations are obtained by equating \eqref{eq:par-c1-2} and
\eqref{eq:par-c1-4}. Furthermore, one can show that the forms for
$A_{ij}$ given in \eqref{eq:Miwa-NSS2} and \eqref{eq:a12-c1} agree,
subject to relations \eqref{eq:vert} and \eqref{rsalbe}.

\subsubsection{Special cases}
The discussion above holds in the generic case, where in particular
the curve defined by the dispersion relation is irreducible.  In fact
the Jacobian of the transformation \eqref{eq:par-c1-2} is given by
\[
 \left|\frac{\partial(o_1,o_2,o_3)}{\partial(\alpha,\beta,\gamma)}\right|
=-4\gamma^2(\alpha-\beta)^2(1+\alpha)^2(1+\beta)^2=4(o_1o_2-o_3^2).
\]
Similarly for \eqref{eq:par-c1-4} we get
\[
 \left|\frac{\partial(o_1,o_2,o_3)}{\partial(a,b,c)}\right|
=(a^2-b^2)(b^2-c^2)(c^2-a^2)=3o_1o_2o_4/(abc).
\]
We need to consider separately the cases where the RHS vanishes.  It
is perhaps not so evident that in the second case the Jacobian also
vanishes when $o_3+o_4=0$, since $o_3+o_4=(a-b)(a-c)(b-c)$.  Among the
various special limits we are only interested in those not included in
Cases 2 or 3.

We now consider these special cases.

\paragraph{$\bullet$ Special case $o_1o_2=o_3^2$, $o_i\neq 0$:}
This is a singular point for the parametrization \eqref{eq:par-c1-2}
in the analysis of Sec.~\ref{S431}.

We use the relation to eliminate $o_2$. Then equation \eqref{eq:c1b}
becomes
\begin{align}
f_{n - 1,m + 1}f_{n + 1,m - 1}\,o_1o_3 - f_{n - 1,m}f_{n +
1,m}\,o_3 (o_1+o_3) - &f_{n,m - 1}f_{n,m + 1}\,o_1(o_1 + o_3)\nonumber \\& +
f_{n,m}^2\, (o_1^2+o_1o_3+o_3^2)=0,\label{eq:s123eq}
\end{align}
and the dispersion relation \eqref{eq:DRc1b}
\[
(o_1 P_j+o_3Q_j)^2=(o_1+o_3)^2,
\]
which is solved by
\[
P_j=-o_3'Q_j+\sigma_j(1+o_3'),\quad \sigma_j^2=1,\quad o_3'=o_3/o_1.
\]
Clearly there can be two kinds of solitons, depending on the chosen sign
$\sigma_j$.

By directly solving \eqref{eq:s123eq} one finds the PWF
\begin{subequations}\label{eq:123res1}
\begin{equation}
e^{\eta_j}=\left(\frac{-o_3'Q_j+\sigma_j(1+o_3')-1}
{-o_3'Q_j+\sigma_j(1+o_3')+1}\right)^{n}
\left(\frac{Q_j-1}{Q_j+1}\right)^{m},
\label{eq:pwf-4.3.4}
\end{equation}
and the phase factor
\begin{equation}
A_{ij}=\frac{Q_i^2-(\tfrac12\sigma_i\sigma_j+\tfrac32)Q_iQ_j+Q_j^2
  -(Q_i-Q_j)(\sigma_i-\sigma_j)(\tfrac12+\tfrac1{o_3'})-
\tfrac12(1-\sigma_i\sigma_j)}
{Q_i^2-(\tfrac12\sigma_i\sigma_j-\tfrac32)Q_iQ_j+Q_j^2
  -(Q_i+Q_j)(\sigma_i+\sigma_j)(\tfrac12+\tfrac1{o_3'})-
\tfrac12(1+\sigma_i\sigma_j)}.
\label{eq:A12-4.3.4}
\end{equation}
\end{subequations}

The results \eqref{eq:123res1} can be obtained from \eqref{eq:solC1b}
with the singular limit $\varepsilon\to0$ using
\begin{eqnarray*}
\alpha&=&-1+2\varepsilon o_1+O(\varepsilon^2),\\
\beta&=& -1-2\varepsilon o_3+O(\varepsilon^2),\\
\gamma&=&-1/(8\varepsilon^2 o_1)+O(\varepsilon^{-1}),
%\\+(o_3-o_1)/(4o_1\varepsilon)+o_1(6o_3-o_1)/(8o_1)+\dots
\end{eqnarray*}
and
\[
k_i=\sigma_i-\varepsilon o_3(Q_i-\sigma_i)+O(\varepsilon^2),
\]
where $\varepsilon$ is defined by
$o_1o_2-o_3^2=-\varepsilon^2o_3^2(o_1+o_3)^2$.

As for the the Miwa's parametrization \eqref{eq:par-c1-4} the relation
$o_1o_2=o_3^2$ is a regular point. For convenience let us use the
parametrization
\[
%a=-(3s+1)(s+1)u,\quad b=(3s-1)(s-1)u,\quad c=(3s+1)(3s-1)su,
a=v(1+s)/(3s-1),\quad b=v(1-s)/(3s+1),\quad c=-sv,
\]
we get
\[
o_1=-2\frac{(3s^2+1)(s+1)^2v^3}{(3s+1)^2},\quad
o_2=-2\frac{(3s^2+1)(s-1)^2v^3}{(3s-1)^2},\quad
o_3=2\frac{(3s^2+1)(s^2-1)v^3}{9s^2-1},
%o_1=(3s-1)^2(s+1)^2v,\quad o_2=(3s+1)^2(s-1)^2v,\quad
%o_3=-(3s^2-1)(s^2-1)v,
\]
where $v,s$ are the new parameters.  Then the reduction condition
\eqref{eq:cond1b} is satisfied by choosing
\[
%p_i=\frac{\sigma\, u}{9s^2-1} \frac{3k_i-1}{k_i+1},\quad
%q_i=\frac{\sigma\, u}{9s^2-1} \frac{3k_i+1}{k_i-1},
p_i=\frac{\sigma_i}{ v} \frac{3\ell_i-1}{\ell_i+1},\quad
q_i=\frac{\sigma_i}{ v}\frac{3\ell_i+1}{\ell_i-1},
\]
which again shows the possibility of two different kinds of solitons,
due to $\sigma_i=\pm1$.

With these choices the phase factor and the PWFs of
\eqref{eq:Miwa-NSS} only depend on
$\ell^2$ and it is then convenient to replace
\begin{equation}\label{eq:tra123}
\ell_i^2=-\frac13\frac{k'_i+4r}{k'_i-4r},\quad
r:=-\frac{2(3s^2+1)^2(s+1)v^3}{(3s+1)^2(3s-1)},
\end{equation}
and with this we get form \eqref{eq:Miwa-NSS} the symmetric formulae
(recalling \eqref{eq:rednm} and dropping the primes)
\begin{subequations}\label{eq:123res2}
\begin{equation}
e^{\eta_j}=\left(\frac{k'_i+(o_1+2o_3)-3\sigma_j o_1}
{k'_i+(o_1+2o_3)+3 \sigma_j o_1}\right)^{n}
\left(\frac{k'_i-(2o_1+o_3)+3 \sigma_j o_3}
{k'_i-(2o_1+o_3)-3 \sigma_j o_3}\right)^{m},
\end{equation}
\begin{equation}
(A_{ij})^{\sigma_i\sigma_j}=\frac{(k'_i-k'_j)^2}
{{k'_1}^2+k'_1k'_2+{k'_2}^2-12(o_1^2+o_1o_3+o_3^2)}.
\end{equation}
\end{subequations}
We note that $r$ solves
\begin{equation}\label{rrdef}
r^2=o_1^2+o_1o_3+o_3^2,
\end{equation}
which appears later as well, and that the transformation \eqref{eq:tra123}
is  singular if $r=0$. Finally the result \eqref{eq:123res2} is related to
\eqref{eq:123res1} via
\[
k'_i=-3\sigma_i o_3 Q_i+2o_1+o_3.
\]

\paragraph{$\bullet$ Special case $o_3+o_4=0$:} (i.e.,
$o_1+o_2+o_3=0$). This is a regular point for the derivation of
Sec.~\ref{S431} but a singular point for Sec.~\ref{S432}.  It leads to
a 3-term equation,
\begin{equation}
  \label{eq:c1b34}
o_3\,f_{n+1,m-1} f_{n-1,m+1} + o_1\,f_{n+1,m} f_{n-1,m} -
(o_1 + o_3) f_{n,m+1} f_{n,m-1} =0,
\end{equation}
which is different from the 3-term equations in Cases 2 and 3.

In the
analysis of Sec.~\ref{S431} we should just parameterize $\alpha,
\beta, \gamma$ by
\[
\alpha=(o_1+r)^2/(o_3(o_1+o_3)),\,
\beta=-(r+o_1+o_3)^2/(o_1o_3),\,
\gamma=-(2 r^2-o_3^2-r (2 o_1+o_3))^2/(8o_3r^2),
\]
where $r^2$ is as before \eqref{rrdef}. If we now reparametrize
$k_j=\kappa_j(2o_1+o_3+2r)/(3o_3)$ then the PWFs \eqref{eq:pwf-c1}
become \footnote{For these computations it is convenient to use the
  rational parametrization $o_1=z(9w^2-1)/(3w^2+1),\,
  o_2=[4(3w-1)wz]/[(3w^2+1)(w-1)],\,
  o_3=-[(3w-1)^2(w+1)z]/[(3w^2+1)(w-1)]$ and correspondingly
  $\alpha=(w + 1)/[(3w - 1)w],\,\beta=[(3w + 1)(w + 1)]/[(3w - 1)(w -
  1)],\, \gamma=z[(3w - 1)^4(w - 1)w^2]/[2(3w^2 + 1)^3(w + 1)]$,
  $r=-(3w-1)z/(w-1)$, where
  $w,z$ are the two new free parameters.}
\begin{subequations}\label{eq:34}
\begin{eqnarray}
e^{\eta_j}&=&
\left[\frac{(o_3\kappa_j+2o_1+o_3-2r)(o_2\kappa_j+2o_1+o_2+2r)}
{(o_3\kappa_j-2o_1-o_3+2r)(o_2\kappa_j-2o_1-o_2-2r)}\nonumber\right]^n\\
&&\times\left[\frac{(o_3\kappa_j+2o_1+o_3-2r)(o_1\kappa_j+2o_3+o_1+2r)}
{(o_3\kappa_j-2o_1-o_3+2r)(o_1\kappa_j-2o_3-o_1-2r)}\right]^m
\end{eqnarray}
and $A_{ij}$ \eqref{eq:a12-c1} factors as
\begin{equation}
A_{ij}=\frac{[(\kappa_i\kappa_j-3)^2-(\kappa_i+\kappa_j)^2](\kappa_i-\kappa_j)^2}
{[(\kappa_i\kappa_j+3)^2-(\kappa_i-\kappa_j)^2](\kappa_i+\kappa_j)^2}.
\end{equation}
\end{subequations}

For the singular limit that should be used with Sec.~\ref{S432} we
start with
\[
o_1=-(o_2+o_3)(1+o_2 o_3\epsilon^6),
\]
and get the expansions\footnote{This is also the type of limit by
  which Miwa's equation reduces to the Hirota's equations.}
\begin{eqnarray*}
%  a&=&\tfrac12\epsilon^{-1}+\tfrac13(o_3-o_2)\epsilon^2+O(\epsilon^5),\\
  a&=&1+\tfrac23(o_3-o_2)\epsilon^3+O(\epsilon^6),\\
%  b&=&\tfrac12\epsilon^{-1}-\tfrac13(o_2+2o_3)\epsilon^2+O(\epsilon^5),\\
  b&=&1-\tfrac23(o_2+2o_3)\epsilon^3+O(\epsilon^6),\\
%  c&=&\tfrac12\epsilon^{-1}+\tfrac13(2o_2+o_3)\epsilon^2+O(\epsilon^5),\\
  c&=&1+\tfrac23(2o_2+o_3)\epsilon^3+O(\epsilon^6),
\end{eqnarray*}
and
\begin{eqnarray*}
  p(\kappa )&=&-\sigma-2\frac{2 r  \kappa  (2 o_1+o_3)[ 3
    (\kappa ^2-1) o_3-4 \kappa      \sigma r]
    +3 o_3 \sigma r (\kappa ^2+1) [(\kappa ^2-1) o_3+4 \kappa  r  \sigma]}
  {3 [16 \kappa ^2 r^2+3 o_3^2 (\kappa ^2-1)^2]}\epsilon^3+O(\epsilon^6),\\
%  p(\kappa )&=&-2\sigma\epsilon-4\frac{2 r  \kappa  (2 o_1+o_3)[ 3
%    (\kappa ^2-1) o_3-4 \kappa      \sigma r]
%    +3 o_3 \sigma r (\kappa ^2+1) [(\kappa ^2-1) o_3+4 \kappa  r  \sigma]}
%  {3 [16 \kappa ^2 r^2+3 o_3^2 (\kappa ^2-1)^2]}\epsilon^4+O(\epsilon^7),\\
  q(\kappa )&=&-p(-\kappa ).
\end{eqnarray*}
Then in the limit $\epsilon \to 0$ one obtains \eqref{eq:34} from
\eqref{eq:Miwa-NSS}.

\paragraph{$\bullet$ Special case
  $o_1=\omega,\,o_2=\omega^2,\,o_3=1,\,\omega^2+\omega+1=0$}
The combination of the previous two cases implies that $o_i$ are cubic
roots of unity. For convenience we choose $o_j=\omega^j$, where
$\omega^3=1,\omega\neq 1$, and then the equation becomes
\[
f_{n-1,m+1}f_{n+1,m-1}+\omega f_{n-1,m}f_{n+1,m}+\omega^2 f_{n,m-1}f_{n,m+1}=0.
\]

For analyzing this case we have to take a singular limit in both
methods. We
approach the limit with $o_1=\omega+\epsilon^6$, $\epsilon\to 0$,
and then we get the expansions ($\varepsilon=i\omega^2\epsilon$)
\begin{eqnarray*}
\alpha&=&-1-2i\omega\epsilon^3+2\omega^2\epsilon^6+O(\epsilon^9),\\
\beta&=&-1+2i\epsilon^3+2\epsilon^6+O(\epsilon^9),\\
\gamma&=&\tfrac18\omega^2\epsilon^{-6}-i\tfrac14(\omega+2)\epsilon^{-3}+O(1)
%\tfrac18(6-\omega)+O(\epsilon^3).
\end{eqnarray*}
and (directly or as limits of the $o_3+o_4=0$ case)
\begin{eqnarray*}
  a&=&1+\tfrac23(\omega+2)\epsilon^3
-\tfrac1{6}(3\omega+4)\epsilon^6+O(\epsilon^9), \\
%  a&=&\tfrac12\epsilon^{-1}+\tfrac13(\omega+2)\epsilon^2
%-\tfrac1{12}(3\omega+4)\epsilon^5+O(\epsilon^8), \\
  b&=&1+\tfrac23(\omega-1)\epsilon^3
-\tfrac5{6}\omega\epsilon^6+O(\epsilon^9), \\
%  b&=&\tfrac12\epsilon^{-1}+\tfrac13(\omega-1)\epsilon^2
%-\tfrac5{12}\omega\epsilon^5+O(\epsilon^8), \\
%  c&=&\tfrac12\epsilon^{-1}-\tfrac13(2\omega+1)\epsilon^2
%-\tfrac1{12}(\omega+4)\epsilon^5+O(\epsilon^8).\\
  c&=&1-\tfrac23(2\omega+1)\epsilon^3
-\tfrac1{6}(\omega+4)\epsilon^6+O(\epsilon^9).
\end{eqnarray*}
Next in order to have nontriial PWFs and relations \eqref{eq:vert} we
must also take singular limits in $k,p,q$, namely
\begin{eqnarray*}
k_j&=&\sigma_j[1+i\tfrac13\epsilon^3(-k'_j+2(\omega-1))],\\
%p_j&=&2\epsilon\sigma'+4\frac{\epsilon^4}{k'_j}
%(\sigma_j(1-\omega)+\sigma'(\omega+1))+O(\epsilon^7),\\
p_j&=&\sigma'+2\frac{\epsilon^3}{k'_j}
(\sigma_j(1-\omega)+\sigma'(\omega+1))+O(\epsilon^6),\\
%q_j&=&-2\epsilon\sigma'+4\frac{\epsilon^4}{k'_j}
%(\sigma_j(1-\omega)-\sigma'(\omega+1))+O(\epsilon^7).\\
q_j&=&-\sigma'+2\frac{\epsilon^3}{k'_j}
(\sigma_j(1-\omega)-\sigma'(\omega+1))+O(\epsilon^6).
\end{eqnarray*}
Here $\sigma_j,\sigma'=\pm1$. In the limit $\epsilon\to 0$ we then get
from \eqref{eq:Miwa-NSS}
\begin{subequations}
\begin{equation}
e^{\eta_j}=\left(\frac
{k'_j\omega^2-2\omega-1-3\sigma_j}
{k'_j\omega^2-2\omega-1+3\sigma_j}
{}\right)^n
\left(\frac
{k'_j-2\omega-1+3\sigma_j}
{k'_j-2\omega-1-3\sigma_j}
\right)^{m}\,,
\end{equation}
and
\begin{equation}
(A_{ij})^{\sigma_i\sigma_j}=\frac{(k'_i-k'_j)^2}
{{k'_i}^2+{k'}_i{k'}_j+{k'_j}^2}\, ,
\end{equation}
\end{subequations}
which also follow directly form \eqref{eq:123res2}. These also follow
as singular limits from \eqref{eq:34}, if we use
$\kappa_j=-\sigma_j(2\omega+1)(1+i\epsilon^3 k_j'/3)$, along with
$o_1=\omega+\epsilon^6,\,o_2=\omega^2-\tfrac47(2\omega+3)\epsilon^6,
\, o_3=-o_1-o_2+O(\epsilon^6)$.

\paragraph{$\bullet$ Special case $o_1=0$, $o_2o_4\neq 0$:}
This can be obtained from the analysis of \ref{S431} by just putting
$\beta=0$ but for \eqref{eq:par-c1-4} a singular limit is needed. Thus
we let
\[
\alpha=o_4/o_2,\,\beta=\epsilon^3\, o_4/(4o_3^2),\,
\gamma=o_2^2/(2o_4),
\]
\[
a=a_m /\epsilon,\,b=-\epsilon^2\,o_2/(2a_m ^2),
c=-\epsilon^2\,o_4/(2a_m ^2),\text{ where }a_m ^3=-2o_3(o_2+o_3).
\]
Then with the parametrization
\[
p_i=\frac{-4(o_2+o_3)o_3}{a_m o_2\epsilon^2\, k_i},\,
q_i=\frac{-(o_4+o_2k_i^2)\epsilon}{2a_m o_3k_i},
\]
we find that in the limit $\epsilon\to 0$ equation \eqref{eq:par-c1-4} is
satisfied and the formulae of Sec.~\ref{S432} work. From both
approaches the result is
\begin{equation}
  \label{eq:pwf-c1-new}
e^{\eta_i}=
  \left(\frac{(k_i-1)(k_i-\alpha)}{(k_i+1)(k_i+\alpha)}\right)^n
  \left(\frac{k_i-1}{k_i+1}\right)^m,
\quad
   A_{12}=\frac{(k_1-k_2)^2\,(k_1 k_2 -  \alpha)}
{(k_1+k_2)^2\,(k_1 k_2 +  \alpha)}.
\end{equation}
The difference equation in this sub-case reads
\begin{equation}
  \label{eq:c1bo1}
(f_{n+1,m-1} f_{n-1,m+1}  - f_{n,m+1} f_{n,m-1})o_3 - (f_{n+1,m} f_{n-1,m}
 - f_{n,m}^2 )(o_3 +o_2)=0.
\end{equation}

The case $o_1=0$, $o_2o_4\neq 0$  is similar.

The further specialization to $o_1=o_4=0$ can be obtained from the
above by letting first $o_2=-2o_3+\rho\epsilon$, i.e.,
$o_4=-\rho\epsilon -\epsilon^3$. This corresponds to $\alpha\to 0$ in
\eqref{eq:pwf-c1-new} and
\begin{equation}
  \label{eq:c1bo1o4}
  f_{n+1,m-1} f_{n-1,m+1}  - f_{n,m+1} f_{n,m-1} + f_{n+1,m} f_{n-1,m}
  - f_{n,m}^2 =0.
\end{equation}

Similarly the case  $o_1=o_2=0$ can be obtained by further putting
$o_2=\epsilon$. This corresponds to $\alpha\to\infty$, which yields
\begin{equation}
  \label{eq:pwf-o12}
  e^{\eta_i}=
  \left(\frac{1-k_i}{1+k_i}\right)^n
  \left(\frac{k_i-1}{k_i+1}\right)^m,
  \quad
  A_{12}=-\left(\frac{k_1-k_2}{k_1+k_2}\right)^2,
\end{equation}
and
\begin{equation}
  \label{eq:c1bo1o2}
  f_{n+1,m-1} f_{n-1,m+1}  - f_{n,m+1} f_{n,m-1} - f_{n+1,m} f_{n-1,m}
  + f_{n,m}^2 =0.
\end{equation}
Note that here $A_{12}$ is negative so this system can have a
non-singular 2SS only if it composed from a complex conjugate pair of
parameters $k_i$. In both of these double zero cases the solitons have
a fixed angle in the $n,m$-space.

\paragraph{$\bullet$ Special case $o_4=0$ $(o_1+o_2+2o_3=0)$:}
This is a regular point for Sec.~\ref{S432} but a singular point for
Sec.~\ref{S431}.  The dispersion relation is
\[
(P-Q)(o_1P -o_2 Q)=0,
\]
and thus we have two kinds of solitons if $o_2\neq o_1$. The
nontrivial factor $(o_1P -o_2 Q)$ is included in the parametrization
\eqref{eq:par-c1-1} provided that $k_i\to 0$ together with $o_4\to 0$.
We will therefore take $\epsilon:=o_4,\, k_i =
k_i'\,\epsilon/(o_1-o_2)^2$ with $k_i'$ arbitrary (the $(o_1-o_2)^2$ factor
is for simplicity later ). For the remaining parameters we
construct a power series expansion in $\epsilon$:
\begin{equation}
\alpha = \frac{o_2}{(o_1-o_2)^2} \epsilon+\dots,\quad
\beta = \frac{o_1}{(o_1-o_2)^2} \epsilon+\dots,\quad
\gamma =  \tfrac12(o_1-o_2)^2/\epsilon+\dots.
%\alpha = \alpha_1 \epsilon+\dots,\quad
%\beta = \beta_1 \epsilon+\dots,\quad
%\gamma =  \gamma_{-1}/\epsilon+\dots
\end{equation}
The limit of the PWF is given by
\begin{equation}
e^{\eta_i}=\left(\frac{o_2-k_i'}{o_2+k_i'}\right)^n
\left(\frac{o_1-k_i'}{o_1+k_i'}\right)^m,
\end{equation}
the phase factor $A_{ij}$ gets the form
\begin{equation}
  A_{ij}=\left(\frac{k_i'-k_j'}{k_i'+k_j'}\right)^2\,
\left(\frac{o_2o_1+k_1'k_2'}{o_2o_1-k_1'k_2'}\right),
\end{equation}
and the bilinear equation
\eqref{eq:c1b} reads
\begin{equation}
(o_2+o_1)f_{n+1,m-1} f_{n-1,m+1} +(o_2-o_1) f_{n+1,m} f_{n-1,m}  -
(o_2-o_1)f_{n,m+1} f_{n,m-1} -(o_2+o_1) f_{n,m}^2 =0.
\end{equation}

In order to keep the connection to the Miwa equation as given
in Equations (\ref{eq:par-c1-4}-\ref{rsalbe}) we need first of
all the expansions for $a,b,c$. The leading terms are
\[
a=a_0+\dots,\quad b=-a_0\frac{o_1}{o_2}+\dots,\quad
c=a_0\frac{o_1}{o_1^2-o_2^2}\epsilon+\dots,\text{ where
}a_0^3=\frac{o_2^2}{2o_1}.
\]
Finally to connect $p_i,q_i$ in (\ref{eq:cond1b}-\ref{eq:vert2}) to
$k_i'$ we find for $p_i,q_i$ the expansions
\[
p_i=\frac{o_2}{a_0\,k_i'}+\dots,\quad
q_i=\frac2{\epsilon}\frac{(o_1 + o_2) (o_1 - o_2)^2 a_0^2\, k_i'}
{({k_i'}^2 - o_1 o_2) o_2^2}+\dots.
\]
It is easy to see that this satisfies \eqref{eq:vert}
but in order to satisfy \eqref{eq:cond1b} is actually necessary to
compute the second order in the $\epsilon$-expansions, while for
\eqref{rsalbe} three orders are necessary.

\paragraph{$\bullet$ Special case $o_4=0$ with others.}
The special case where $o_4=0$ together with $o_2=o_1$, or $o_1=0$, or
$o_2=0$ leads to dispersion relations $Q(P-Q)=0$ and $P(P-Q)=0$,
respectively, which are perhaps too simple to be of interest.

\section{Conclusions}
Hirota's direct method has turned out to be very efficient in
deriving soliton solutions for a given equation.
It has also be used as a method for searching for integrable
equations in the continuous case, the criterion being the existence of
three-soliton solutions.

Here we have applied the three-soliton condition to two-dimensional
partial difference equations. In comparison with the continuous case
there is now much more freedom, since the same partial derivative can
be discretized in many ways. In this paper we restricted the search to
partial-difference equations defined on a $3\times 3$ stencil. For the
$2\times N$ case, see \cite{RIAM}.

The master bilinear \PdE s are Hirota's three-term DAGTE equation
\eqref{eq:dcDAGTE} and Miwa's four term equation \eqref{eq:HM}. In the
detailed analysis of the results found here we showed that all of them
can be obtained from either one of these master equations with a
suitable projection. However, sometimes a nontrivial singular
parameter limit was also needed. Since the phase factor of the soliton
solution is the same for a discrete equation and its continuum limit
we can say that most of the results can be considered as discrete
versions of the KdV equations, but some are related to the Boussinesq
or Ito equations, while some cannot yet be definitely identified.

\subsection*{Acknowledgments}
This project is partially supported by the NSF of China
(No. 11071157), SRF of the DPHE of China (No. 20113108110002) and
Shanghai Leading Academic Discipline Project (No. J50101).

\end{document}